\documentstyle[11pt,newpasp,twoside,epsf]{article}
\def\edcomment#1{\iffalse\marginpar{\raggedright\sl#1\/}\else\relax\fi}
\reversemarginpar

\begin{document}
\title{A Sample of X-Ray active extragalactic Sources suitable for NIR
Adaptive Optics Observations}
\author{Jens Zuther, Andreas Eckart, Christian Straubmeier}
\affil{1. Physikalisches Institut, Universit\"at zu K\"oln, Z\"ulpicher
Str. 77, 50937 K\"oln, Germany}
%\author{Andreas Eckart}
%\affil{1. Physikalisches Institut, Universit\"at zu K\"oln, Z\"ulpicher
%Str. 77, 50937 K\"oln, Germany}
\author{Wolfgang Voges}
\affil{Max-Planck Institut f\"ur extraterrestrische Physik, 85748
Garching, Germany}
%\author{Christian Straubmeier}
%\affil{1. Physikalisches Institut, Universit\"at zu K\"oln, Z\"ulpicher
%Str. 77, 50937 K\"oln, Germany}

\begin{abstract}
Recent X-ray surveys have now resolved most of the X-ray background (XRB)
into discrete sources. While this represents a breakthrough in the
understanding of the XRB, the astrophysical nature of these sources
still remains mysterious. In this article we present a sample of
X-ray/optically selected 
extragalactic objects which are suitable for adaptive optics
observations in the near infrared (NIR) at highest angular
resolution. The sample is based on a cross-correlation of the Sloan
Digital Sky Survey and the ROSAT All Sky Survey. The NIR properties
can help to disentangle the nature of the X-ray bright, partially
absorbed and spectroscopically passive background objects and their
hosts.

\end{abstract}
\section{Introduction}
\subsection{Near Infrared Adaptive Optics}
Adaptive Optics (AO) systems on large telescopes like NACO at the VLT
(Brandner et al. 2002) overcome the limitations introduced by earth's
turbulent atmosphere in terms of image degradation and allow 
imaging and spectroscopy at the diffraction limit of these
telescopes (see Beckers 1993 for a review). For example an 8m-class
telescope offers an angular resolution of $\sim$50~mas at
1.65~$\mu$m. AO therefore enables the study of extragalactic targets 
at highest spatial resolution and e.g. directly allows the comparison
of more distant galaxies with nearby ones, observed without AO in the
same wavelength domain, at the same spatial resolution. 

The near infrared (NIR) is a sensitive tracer of the mass dominating
(older) stellar populations in galaxies. At the same time the NIR is
less affected by extinction, but still sensitive to the
distribution and contribution of (warm) dust.
Studying the detailed morphology, dynamics and composition of the
sources described below is therefore ideally done in the NIR since
especially the circum-nuclear regions of the targets are expected to
be extincted.

\subsection{The X-Ray Background}
The X-ray background (XRB) in the 0.5-10~keV regime has been resolved
into discrete sources by recent work on X-ray surveys like ROSAT,
Chandra, and XMM (e.g. Hasinger 1998; Miyaji, Hasinger, \& Schmidt
2000; Mushotzki et al. 2000; Giacconi et al. 2001; Brandt et
al. 2001b). However, the astrophysical nature of these sources still
remains unknown. Especially two subjects are of interest in this
context. \textbf{The hardness of the X-ray spectra:} It is found that
the cosmic XRB is much harder than the X-ray emission of unobscured
(type I) AGN in the local universe. Therefore the existence of a
substantial obscured AGN (type II) population is required. These are
found preferentially at lower redshifts ($z\sim 0.6$), in contrast to
predictions of XRB models (Gilli 2003 and references therein). An
important question in understanding the nature of sources with hard
X-ray spectra is whether they are hardened due to a large amount of
intrinsic (circum-nuclear) absorption or whether the spectrum of the
central engine itself is intrinsically hard. It is quite likely that
the X-ray spectra are of composite nature, i.e. both the emission and
absorption mechanisms are of importance. In this case it has to be
determined how the nuclear properties are correlated with the
properties of the corresponding hosts. \textbf{Evolution of X-ray 
active AGN:} A recent result (Tacconi et al. 2002) is that the more
(spectroscopically) passive X-ray-bright, early-type galaxies may
originate from a population of ULIRG galaxies that contain  QSO-like
active galactic nuclei. Some of the ULIRGs resemble local QSOs in
their NIR and bolometric luminosities because they are very
efficiently transforming dust and gas into stars and/or feed their
central engine. However, ULIRGs have smaller effective radii and
velocity dispersions than the local QSO/radio galaxy population. This
then implies that their corresponding host and  black hole masses 
must be smaller. Smaller Black holes are found in local Seyfert
galaxies. Indeed, based on optical SDSS spectra, the sources in
our sample resemble those of LINERs or Seyferts (Fig. 1). It is therefore
likely that they do not evolve into optically bright QSOs but rather
into quiescent field ellipticals or in a still active state into
X-ray-bright, early-type galaxies that can be found in ROSAT and
CHANDRA based samples. If this finding is correct then our X-ray based
sample should reveal host galaxy properties 
%($\sigma$, $r_{eff}$,$\mu_{eff}$, M$_K$) 
quite similar to AGN-dominated and star formation
dominated ULIRGs. In other words they should fall close to the
fundamental plane of early-type galaxies. 
The question is: How do they compare to L$_*$ rotating ellipticals,
giant ellipticals, optically/UV-bright, and low-z QSOs/radio galaxies,
i.e. what is their parent population?

\begin{figure}[ht]
\plotone{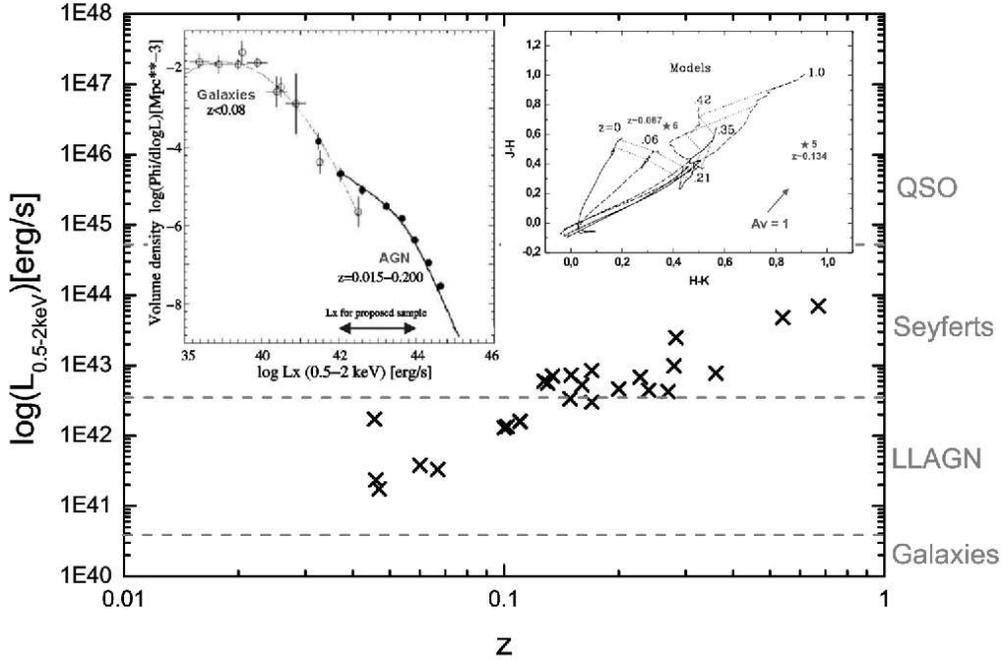}
\caption{Hubble diagram of the sources of the 'initial'
sample. Different source types are indicated on the right axis
 (Hasinger, private communication). \emph{Left inset} displaying the
luminosity function of AGN compared to those of normal
galaxies(Hasinger, private communication). \emph{Right inset}
comparing NIR-colors of sources of the sample (stars) with those of
Gissel-stellar-evolutionary model colors (Hutchings \& Neff 1997).} 
\end{figure}

\section{The ROSAT/SDSS based Sample}
The Sloan Digital Sky Survey (SDSS, e.g. York et al. 2000) with its
multicolor photometric and spectroscopic data presents a rich source
to search for astrophysically interesting objects. 

To address the above stated questions, using a cross-correlation of
SDSS optical and ROSAT X-ray data provided by the Early Data Release
(EDR, Stoughton et al. 2002) and the ROSAT All Sky Survey (RASS, Voges
et al. 1999), we searched for extragalactic, X-ray active objects
brighter than $r < 20$ with a natural guide star (NGS) brighter
than $r<15$ in their vicinity. These magnitude units provide
reasonable AO performance on 8m-class telescopes. Recent AO
implementations allow NGS/science target angular separations up to
$40''$. These criteria resulted in a set of 27 galaxy/NGS pairs. 

To only select the best optical X-ray counterparts we rejected those
objects fulfilling one ore more of the following conditions:\\
(1) \emph{Angular separation between optical and X-ray position
larger than $40''$.} The SDSS database identifies ROSAT sources with
galaxies/AGN if the separation between both positions is less than
$60''$.\\
(2) \emph{Hardness ratio of -1 or a stellar optical color
$u-g<1.3$.} These are indicative for X-ray active stars, the former
typical for white dwarfs. In this case the X-ray emission might be
dominated by the NGS rather than the galaxy. \\
(3) \emph{X-ray emission which shows excess extend over the ROSAT
psf width.} Here the X-ray flux is likely to be dominated by emission
from an extended hot gas component at the center of a galaxy cluster
(cf. Voges et al. 1999).

The final sample consists of 12 galaxy/NGS pairs in the redshift range
$0.06<z<0.67$. Their X-ray luminosities (calculated, assuming a
power-law index of -2 and an average galactic absorption column
density of $3\times 10^{20}$cm$^{-2}$) range from
$10^{42}$~erg~s$^{-1}$ to $10^{43}$~erg~s$^{-1}$ (Fig. 1).
Of these sources five objects have SDSS spectra, one of which appears
as a 'passive' elliptical, although its X-ray luminosity indicates
nuclear activity (cf. Comastri et al. 2002). For the remaining
galaxies we used their photometric redshifts. Another source also
shows a red nucleus (source no. 5 in right inset of Fig. 1) contained
in the 2Mass survey, indicating the presence of warm dust, heated by
star formation and/or the central engine.

\section{Conclusions}
The SDSS and its cross-correlations with other surveys provide means
to find suitable extragalactic sources for follow-up NIR observations
at highest angular resolution and sensitivity of 8m-class
telescopes. Large telescopes are also necessary, because the optical
counterparts of deep X-ray sources become very faint. NIR observations
give complementary information to the optical and X-ray data in terms
of highest resolution images of the detailed structure of the host
galaxies. NIR colors and spectra give information on the stellar
content of the host galaxies and the importance of the presence of
dust. Furthermore dynamical information from spectra give estimates for
host masses and possibly black hole masses, since the NIR emission is a
more accurate tracer of the mass in galaxies.

Our sample represents a significant first step to a statistically
relevant sample of X-ray active extragalactic sources observable with
adaptive optics on large telescopes like the VLT. 

\end{document}